\documentstyle[11pt]{article}

\begin{document}
\title{Expansion of the Yang-Mills Hamiltonian in spatial derivatives and glueball spectrum}
\author{Hans-Peter Pavel \\[1cm]
Bogoliubov Laboratory of Theoretical Physics,
\\
Joint Institute for Nuclear Research, Dubna, Russia\\
and\\
Institut f\"ur Kernphysik, Technische Universit\"at Darmstadt
\\
D-64289 Darmstadt, Germany\footnote{email: hans-peter.pavel@physik.tu-darmstadt.de}}
\date{Januar 19, 2010}
\maketitle

\begin{abstract}
A strong coupling expansion of the $SU(2)$ Yang-Mills quantum Hamiltonian
is carried out in the form of an expansion in the number of spatial derivatives,
using the symmetric gauge $\epsilon_{ijk}\! A_{jk}=0$.\newline
Introducing an infinite lattice with box length $a$, I obtain
a systematic strong coupling expansion of the Hamiltonian in $\lambda\equiv g^{-2/3}$,
with the free part being the sum of Hamiltonians of Yang-Mills quantum mechanics
of constant fields for each box, and interaction terms of higher and higher number
of spatial derivatives connecting different boxes.
The corresponding deviation from the free glueball spectrum,
obtained earlier for the case of the Yang-Mills quantum mechanics of spatially constant
fields, is calculated using perturbation theory in $\lambda$.
As a first step, the interacting glueball vacuum and the energy spectrum of the
interacting spin-0 glueball are obtained to order $\lambda^2$. Its relation to the renormalisation
of the coupling constant in the IR is discussed, indicating the absence of infrared fixed points.
\end{abstract}

%Keywords: Quantum field theory, glueball spectrum, lattice-QCD, Hamiltonian formulation.
%%%%%%%%%%%%%%%%%%%%%%%%%%%%%%%%%%%%%%%%%%%%%%%
\section{Introduction}

A very promising method for non-perturbative investigations of Yang-Mills theory has turned
out to be the Hamiltonian approach \cite{Christ and Lee}, in particular the possibility
to use the powerful variational method.

I shall consider here the Yang-Mills theory of \(SU(2)\) gauge fields \( A^a_{\mu}(x) \),
defined by the  action
\begin{equation}
{\cal S} [A] : = - \frac{1}{4}\ \int d^4x\ F^a_{\mu\nu} F^{a\mu \nu}~,\quad
F^a_{\mu\nu} : = \partial_\mu A_\nu^a  -  \partial_\nu A_\mu^a
+ g \epsilon^{abc} A_\mu^b A_\nu^c~,\nonumber
\end{equation}
invariant under both Poincar\'{e} and scale transformations,
and the local $SU(2)$ gauge transformations
$U[\omega(x)]\equiv \exp(i\omega_a\tau_a/2)$
\begin{equation}
A_{a\mu}^{\omega}(x) \tau_a/2   =
U[\omega(x)] \left(A_{a\mu}(x) \tau_a/2 +
{i\over g}\partial_\mu \right) U^{-1}[\omega(x)]~.\nonumber
\end{equation}

The transition to the corresponding quantum theory is then carried out
by exploiting the time dependence of the gauge transformations to put
%\begin{equation}
\begin{eqnarray}
A_{a0}(x) = 0~,\quad a=1,2,3~,\nonumber \quad\quad (\rm{Weyl}\  \rm{gauge})\nonumber
\end{eqnarray}
%\end{equation}
and impose canonical commutation relations on the spatial fields using the Schr\"odinger representation
%\begin{equation}
\begin{eqnarray}
 [\Pi_{ai}({\mathbf{x}}),A_{bj}({\mathbf{y}})]=i\delta_{ab}\delta_{ij}\delta({\mathbf{x}}-{\mathbf{y}})\quad \longrightarrow
\quad \Pi_{ai}({\mathbf{x}})=-E_{ai}({\mathbf{x}})=-i\delta/\delta A_{ai}({\mathbf{x}})~.\nonumber
\end{eqnarray}
%\end{equation}
The physical states $\Psi$ have to satisfy the system of equations
\begin{eqnarray}
&&(H-E)\Psi = 0\quad ({\rm Schr\ddot{o}dinger}\ {\rm equation})~,\nonumber\\
&&G_a({\mathbf{x}})\Psi = 0\quad ({\rm Gauss}\ {\rm law}\ {\rm constraints})~,\label{System}
\end{eqnarray}
with the Hamiltonian
$ ( \  B_{ai}(A) =\epsilon_{ijk}\left(\partial_j A_{ak}+{1\over 2}g
\epsilon_{abc}A_{bj} A_{ck}\right)\ )$
\begin{equation}
H = \int d^3{\mathbf{x}} {1\over 2}
\sum_{a,i} \left[\left(\frac{\delta}{\delta A_{ai}({\mathbf{x}})}\right)^2+B_{ai}^2(A({\mathbf{x}}))\right]~,
\nonumber
\end{equation}
 and the Gauss law operators
\begin{equation}
G_a ({\mathbf{x}}) = -i\left(\delta_{ac}\partial_i +g\epsilon_{abc}A_{bi}({\mathbf{x}})\right)
             \frac{\delta}{\delta A_{ci}({\mathbf{x}})}~,\quad a=1,2,3~,\nonumber
\end{equation}
which are the generators of the residual time-independent gauge transformations,
commute with the Hamiltonian and satisfy angular momentum commutation relations
%\begin{equation}
\begin{eqnarray}
[G_a({\mathbf{x}}),H]=0~,\quad\quad [G_a({\mathbf{x}}),G_b({\mathbf{y}})]=
ig\epsilon_{abc}G_c({\mathbf{x}})\delta({\mathbf{x}}-{\mathbf{y}})~,\nonumber
\end{eqnarray}
%\end{equation}
The matrix elements are
\begin{equation}
\langle\Phi_1|{\cal O} |\Phi_2\rangle=
\int \prod_{ik} dA_{ik} \ \Phi_1^* {\cal O}\Phi_2~.\nonumber
\end{equation}

\section{Physical $SU(2)$ Quantum Hamiltonian in the symmetric gauge}

In order to calculate the eigenstates and their energies, it is useful to implement
the non-Abelian Gauss law constraints into the Schr\"odinger equation
by further fixing the gauge using the remaining time-independent gauge transformations.
One possibility, well suited for the high energy sector of the theory, is to
impose the Coulomb gauge $\chi_a(A)=\partial_i A_{ai}=0$ describing the dynamics in terms
of physical colored transverse gluons.
I shall here choose the symmetric gauge\cite{KP1}\footnote{
%------------------------------------------------------------------------------------
It has been proven in \cite{KMPR}, that the symmetric gauge exists (at least for strong coupling),
by showing that any time-independent gauge field
can be carried over uniquely into the symmetric gauge.
%------------------------------------------------------------------------------------
}
\begin{equation}
\quad\quad\chi_i(A)=\epsilon_{ijk}A_{jk}=0\quad(\rm{symmetric}\ \rm{gauge})~.
\end{equation}
In contrast to the Coulomb gauge, the symmetric gauge allows for
an expansion of the physical Hamiltonian in spatial derivatives, which
makes it very suited for the study of the infrared sector of Yang-Mills theory.
The physical degrees of freedom in the symmetric gauge are the six components
of a colorless local symmetric tensor field.

The symmetric gauge corresponds to the
point transformation to the new set of adapted coordinates,
the three $ q_j\ \ (j=1,2,3)$  and the six elements
$S_{ik}= S_{ki}\ \ (i,k=1,2,3)$ of the positive definite\footnote{
%-------------------------------------------------------------------------------
In the infinite coupling limit this transformation reduces to the polar decomposition,
in which the symmetric matrix can be chosen to be positive definite.
%------------------------------------------------------------------------------
}
symmetric $3\times 3$ matrix $S$
\begin{equation}
\label{coordtrafo}
A_{ai} \left(q, S \right) =
O_{ak}\left(q\right) S_{ki}
- {1\over 2g}\epsilon_{abc} \left( O\left(q\right)
\partial_i O^T\left(q\right)\right)_{bc}\,,\nonumber
\end{equation}
where \( O(q) \) is an orthogonal $3\times 3$ matrix parametrized by the $q_i$.
After the above coordinate transformation (\ref{coordtrafo}), the non-Abelian
Gauss law constraints become the Abelian conditions
\begin{eqnarray}
 G_a\Phi=0\quad \Leftrightarrow\quad\frac{\delta }{\delta q_i}\Phi=0\quad
 (\rm{Abelianisation}),
 \nonumber
\end{eqnarray}
that the physical states should depend only on the physical variables $S_{ik}$, and the system
(\ref{System})
reduces to the unconstrained Schr\"odinger equation
\begin{equation}
H(S,P)\Phi(S)=E\Phi(S)~.
\end{equation}

The correctly ordered physical quantum Hamiltonian \cite{Christ and Lee} in the symmetric gauge
in terms of the physical variables $S_{ik}({\mathbf{x}})$  and the corresponding canonically
conjugate momenta
$P_{ik}({\mathbf{x}})\equiv -i\delta /\delta S_{ik}({\mathbf{x}})$  reads
\begin{eqnarray}
H(S,P)\!\!\!\!&=&\!\!\!\! {1\over 2}{\cal J}^{-1}\!\!\int d^3{\mathbf{x}}\int d^3{\mathbf{y}}
\ P_{mn}({\mathbf{x}}) {\cal J}{\cal K}_{mn|st}({\mathbf{x}},{\mathbf{y}}) P_{st}({\mathbf{y}})
     +{1\over 2}\int d^3{\mathbf{x}}\left(B_{ai}(S)\right)^2~,
\end{eqnarray}
with the kernel
\begin{equation}
\label{kernel}
{\cal K}_{mn|st}({\mathbf{x}},{\mathbf{y}}):=
            \delta_{ms}\delta_{nt}\delta({\mathbf{x}}-{\mathbf{y}})
            -2\langle {\mathbf{x}}\ n|D_m(S)\
      ^{\ast}\! D^{-2}(S)\ D_s(S)|{\mathbf{y}}\ t\rangle~,
\end{equation}
the Jacobian
\begin{equation}
\label{J}
{\cal J}\equiv\det |^{\ast}\! D|~,
\end{equation}
the covariant derivative
\begin{eqnarray}
D_i(S)_{kl}\equiv\delta_{kl}\partial_i-g\epsilon_{klm}S_{mi}~,
\nonumber
\end{eqnarray}
the Faddeev-Popov (FP) operator
\begin{equation}
\label{FP}
^{\ast}\! D_{kl}(S)\equiv \epsilon_{kmi}D_i(S)_{ml}
   =\epsilon_{kli}\partial_i-g\gamma_{kl}(S)~,
   \quad\quad \gamma_{kl}(S)\equiv S_{kl}-\delta_{kl} {\rm tr} S~,
\end{equation}
and the Green function
\begin{eqnarray}
\langle {\mathbf{x}}\ a|^{\ast}\! D^{-2}(S)|{\mathbf{y}}\ b\rangle
\equiv ^{\ast}\!\! D^{-2}_{ab}(S)^{({\mathbf{x}})}[\delta({\mathbf{x}}-{\mathbf{y}})]~.
\nonumber
\end{eqnarray}
The matrix element of a physical operator O is given by
\begin{equation}
\label{ME}
\langle \Psi'| O|\Psi\rangle\
\propto
\int \prod_{\mathbf{x}}\Big[dS({\mathbf{x}})\Big]
{\cal J}\Psi'^*[S] O\Psi[S]~.
\end{equation}

A great advantage of the symmetric gauge - in contrast for example to the Coulomb gauge-,
is that the corresponding FP operator, and hence the non-local terms of the physical
Hamiltonian, can be expanded in the number of spatial derivatives.
The Green function
\begin{eqnarray}
\langle {\mathbf{x}}\ k|^{\ast}\! D^{-1}(S)|{\mathbf{y}}\ l\rangle
\equiv ^{\ast}\!\! D^{-1}_{kl}(S)^{({\mathbf{x}})}[\delta({\mathbf{x}}-{\mathbf{y}})]~,
\nonumber
\end{eqnarray}
corresponding to the FP operator (\ref{FP}),
can be expanded in the number of spatial derivatives
\begin{eqnarray}
\langle {\mathbf{x}}\ k|^{\ast}\! D^{-1}(S)|{\mathbf{y}}\ l\rangle &=&
 -{1\over g}\gamma^{-1}_{k l}({\mathbf{x}})\delta({\mathbf{x}}-{\mathbf{y}})
+{1 \over g^2}\gamma^{-1}_{ka}({\mathbf{x}})\epsilon_{abc}
     \partial_c^{({\mathbf{x}})}
     \left[\gamma^{-1}_{bl}({\mathbf{x}})\delta({\mathbf{x}}-{\mathbf{y}})\right]
\nonumber\\
&&\quad\quad\quad\quad
-{1 \over g^3}\gamma^{-1}_{ka}({\mathbf{x}})\epsilon_{abc}
     \partial_c^{({\mathbf{x}})}\!\!\left[\gamma^{-1}_{bi}({\mathbf{x}})\epsilon_{ijs}
     \partial_s^{({\mathbf{x}})}\!\!
     \left[\gamma^{-1}_{jl}({\mathbf{x}})\delta({\mathbf{x}}-{\mathbf{y}})\right]\right]
+...~.
\nonumber
\end{eqnarray}

\section{Expansion of the Hamiltonian in spatial derivatives}

In order to perform a consistent expansion of the physical Hamiltonian in spatial derivatives,
also the non-locality in the Jacobian ${\cal J}$
has to be taken into account. This will be achieved in the following way.

Writing the FP operator in the form
\begin{eqnarray}
^{\ast}\! D_{kl}(S)\equiv
    =-g\gamma_{km}(S)\Big[\delta_{ml}-{1\over g}\gamma^{-1}_{mn}(S)\epsilon_{nli}\partial_i\Big]
    \equiv -g\gamma_{km}(S)^{\ast}\! \widetilde{D}_{ml}(S)~,
\nonumber
\end{eqnarray}
the Jacobian ${\cal J}$ factorizes
\begin{equation}
{\cal J}={\cal J}_0 \widetilde{\cal J}~,
\end{equation}
with the local
\begin{equation}
{\cal J}_0 \equiv\det |\gamma|=\prod_{\mathbf{x}}\det|\gamma({\mathbf{x}})|~, \quad
\det|\gamma({\mathbf{x}})|=\prod_{i<j}\left(\phi_i({\mathbf{x}})+\phi_j({\mathbf{x}})\right)~,
\quad (\phi_i = {\rm eigenvalues\ of}\ S)
\end{equation}
and the non-local
%\begin{equation}
$\widetilde{\cal J}\equiv\det |^{\ast}\! \widetilde{D}|~.$
%\end{equation}
Now I include the non-local part of the measure into the wave functional
\begin{eqnarray}
\widetilde{\Psi}(S):=\widetilde{\cal J}^{-1/2}\Psi(S)~,
\nonumber
\end{eqnarray}
leading to the corresponding transformed Hamiltonian
$
\widetilde{H}:=\widetilde{\cal J}^{1/2} H \widetilde{\cal J}^{-1/2}~,
$
being Hermitean with respect to the local measure ${\cal J}_0$
\begin{eqnarray}
\label{Htilde}
\widetilde{H}(S,P)\!\!\!\!&=&
    \!\!\!\! {1\over 2}{\cal J}_0^{-1}\!\!\int d^3{\mathbf{x}}\int d^3{\mathbf{y}}
\ P_{mn}({\mathbf{x}})\ {\cal J}_0 {\cal K}_{mn|st}({\mathbf{x}},{\mathbf{y}}) P_{st}({\mathbf{y}})
     +{1\over 2}\int d^3{\mathbf{x}}\left(B_{ai}(S)\right)^2+V_{\rm meas}(S)~,
\end{eqnarray}
on the cost of extra terms \footnote{
%----------------------------------------------------------
Although in principle, $V_{\rm meas}$ is part of the electric term of the
Hamiltonian, I shall treat it separately in this work as "measure term".
%----------------------------------------------------------
} $V_{\rm meas}$ from the non-local factor $\widetilde{\cal J}$
of the original measure ${\cal J}$
\begin{eqnarray}
\!\!\!\!\!\!\!\!\!\!\!\!
V_{\rm meas}(S)\!\!\!\!&=&\!\!\!\! {1\over 4}{\cal J}_0^{-1}\!\!\int\!\! d^3{\mathbf{x}}
                       \frac{\delta}{\delta S_{mn}({\mathbf{x}})}\!\!\left[{\cal J}_0\!\!\int\!\! d^3{\mathbf{y}}
                       {\cal K}_{mn|st}({\mathbf{x}},{\mathbf{y}})
                       \widetilde{\Delta}_{st}({\mathbf{y}})\right]
\!+\!{1\over 8}\!\int\!\! d^3{\mathbf{x}}\!\!\int\!\! d^3{\mathbf{y}}
   \widetilde{\Delta}_{mn}({\mathbf{x}}){\cal K}_{mn|st}({\mathbf{x}},{\mathbf{y}})
   \widetilde{\Delta}_{st}({\mathbf{y}}),
\end{eqnarray}
with the original kernel ${\cal K}$ in (\ref{kernel}) and
\begin{eqnarray}
\!\!\!\!\!\!\!\!\!\!\!\!\!
\widetilde{\Delta}_{mn}({\mathbf{x}})\!\!\!\!\!&:=&\!\!\!\!\!
                    \frac{\delta \ln \widetilde{\cal J}}{\delta S_{mn}({\mathbf{x}})}=
          -g\left(\langle {\mathbf{x}}\ m| ^{\ast}\!D^{-1}|{\mathbf{x}}\ n\rangle-
                  \delta_{mn}\langle {\mathbf{x}}\ k| ^{\ast}\!D^{-1}|{\mathbf{x}}\ k\rangle\right)
                 -\delta({\mathbf{0}})\!\left(\gamma^{-1}_{mn}({\mathbf{x}})-
                          \delta_{mn}\mbox{tr}\gamma^{-1}({\mathbf{x}})\right)~.
\end{eqnarray}
The matrix element (\ref{ME}) of a physical operator $O$ becomes the product of local matrix elements
\begin{equation}
\langle \Psi'| O|\Psi\rangle\
\propto
\int \prod_{\mathbf{x}}\Big[dS({\mathbf{x}})
\prod_{i<j}\left(\phi_i({\mathbf{x}})+\phi_j({\mathbf{x}})\right)\Big]
\Psi'^*[S] O\Psi[S]~.
\end{equation}
The transformed physical Hamiltonian (\ref{Htilde}) can be expanded in the
number of spatial derivatives
\begin{equation}
\tilde{H} = H_0 + \sum_\alpha V^{(\partial)}_\alpha+
\left(\sum_\beta V^{(\Delta)}_\beta+\sum_\gamma V^{(\partial\partial\neq\Delta)}_\gamma\right)
+ ...~,
\end{equation}
with the free part $H_0$ containing no spatial derivatives, the interaction parts
$V^{(\partial)}_\alpha$ containing one spatial derivative, and
$V^{(\Delta)}_\beta,V^{(\partial\partial\neq\Delta)}_\gamma$ containing two spatial derivatives,
and so on.

\subsection{The free part $H_0$}

The free part $H_0$ containing no spatial derivatives reads
\begin{eqnarray}
\!\!\!\!\!\!\!\! H_0 \!\!\!\! &=&\!\!\!\! \int\!\! d^3{\mathbf{x}}{1\over 2}\Bigg[\!\!
\left(P_{mn}\right)^2
-i\delta({\mathbf{0}})\!\!\left[\gamma^{-1}_{mn}(S)-\delta_{mn}\mbox{tr}(\gamma^{-1}(S))\right]\!\!P_{mn}
+{1\over 2}\gamma^{-2}_{mn} {\cal S}_m^{\rm spin} {\cal S}_n^{\rm spin}
+{g^2\over 2}\left(\mbox{tr}^2S^2-\mbox{tr}S^4\right)\!\!
\Bigg],
\end{eqnarray}
with the spin densities
$\label{SYP}
{\cal S}_i^{\rm spin}= 2\epsilon_{ijk}S_{ja}P_{ak}~, i=1,2,3$
(note the factor $2$).

In order to achieve a more transparent form for the reduced Yang-Mills
system
I shall limit myself in this work to the principle orbit configurations
\begin{equation}
\label{range}
0<\phi_1<\phi_2<\phi_3<\infty~,
\end{equation}
for the eigenvalues $\phi_1,\phi_2,\phi_3>0$ of the positive definite symmetric matrix $S$
(not considering singular orbits where two or more eigenvalues coincide)
and perform a principal-axes transformation
\begin{equation}
\label{patransf}
S  =  R(\alpha,\beta,\gamma)\ \mbox{diag}\ ( \phi_1 , \phi_2 , \phi_3 ) \
R^{T}(\alpha,\beta,\gamma)~,
\end{equation}
with the \( SO(3)\) matrix  \({R}\) parametrized by the three Euler
angles $\chi\equiv(\alpha,\beta,\gamma)$.
%%%%%%%%%%%%%%%%%%%%%%%%%%%%%%%%%%%%%%%%%%%%%%%%%%%%%%%%%%%%%%%%%%%%%%%%%%%%
The Jacobian of  (\ref{patransf}) is
$|\partial S/\partial(\alpha,\beta,\gamma,\phi)| \propto
\sin\beta \prod_{i<j}\left(\phi_i- \phi_j\right)$.
The original physical variables can then be written in terms of the new canonical variables as
(using Clebsch-Gordan coefficients)\footnote{
%------------------------------------------------------------------------------
For spin-1 fields $S_i^{(1)}$ I use the Cartesian combinations
$S_{1}^{(1)}\equiv S_{1-}^{(1)}:= (S_{+1}^{(1)}-S_{-1}^{(1)})/\sqrt{2}
                                                \equiv e_1^\alpha S_{\alpha}^{(1)},\
S_{2}^{(1)}\equiv S_{1+}^{(1)}:= i(S_{+1}^{(1)}+S_{-1}^{(1)})/\sqrt{2}
                                                 \equiv e_2^\alpha S_{\alpha}^{(1)},\
S_{3}^{(1)}\equiv S_{0}^{(1)}\equiv e_3^\alpha S_{\alpha}^{(1)}$,
such that e.g. $C_{1i\ 1k}^{2 A}:= e_i^\alpha e_j^\beta C_{1\alpha\ 1\beta}^{2 A}$,
For spin-2 fields $S_A^{(2)}$ I use correspondingly the real combinations
$S_{2+}^{(2)}:=(S_{+2}^{(2)}+S_{-2}^{(2)})/\sqrt{2},\
 S_{2-}^{(2)}:=-i(S_{+2}^{(2)}-S_{-2}^{(2)})/\sqrt{2},\
 S_{1+}^{(2)}:=i(S_{+1}^{(2)}+S_{-1}^{(2)})/\sqrt{2},\
 S_{1-}^{(2)}:=-(S_{+1}^{(2)}-S_{-1}^{(2)})/\sqrt{2}$.
%------------------------------------------------------------------------------
}
\begin{equation}
S_{ik}=C_{1i\ 1k}^{2 A} \left(\phi_3\right)^{\!(2)}_{A}
         +\frac{1}{\sqrt{3}}\delta_{ik}\left(\phi_3\right)^{\!(0)}~,\quad\quad
P_{ik}\!\!=\! C_{1i\ 1k}^{2 A}\! \left(\pi_3,\xi_3\right)^{\!(2)}_{A}\!\!
    + \frac{1}{\sqrt{3}}\delta_{ik}\left(\pi_3\right)^{\!(0)}
\end{equation}
with the spin-0 and spin-2 fields (using Wigner D-functions)
\begin{eqnarray}
\left(\phi_3\right)^{\!(0)} &:=&\left(\phi_1+\phi_2+\phi_3\right)/\sqrt{3}
\\
\left(\phi_3\right)^{\!(2)}_{A} &:=&
\sqrt{\frac{2}{3}}\left[
\left(\phi_3-\frac{1}{2}\left(\phi_1+\phi_2\right)\right)D^{(2)}_{A0}(\chi)
+\frac{\sqrt{3}}{2}\left(\phi_1-\phi_2\right)
D^{(2)}_{A2+}(\chi)
\right]~,
\end{eqnarray}
and (using $\pi_i \equiv -i\delta/\delta \phi_{i}$)
\begin{eqnarray}
\label{new-mom1}
\left(\pi_3\right)^{\!(0)} &:=&\left(\pi_1+\pi_2+\pi_3\right)/\sqrt{3}\\
\label{new-mom2}
\left(\pi_3,\xi_3\right)^{\!(2)}_{A} &:=&
\sqrt{\frac{2}{3}}\left[
\left(\pi_3-\frac{1}{2}\left(\pi_1+\pi_2\right)\right)D^{(2)}_{A 0}(\chi)
+\frac{\sqrt{3}}{2}\left(\pi_1-\pi_2\right) D^{(2)}_{A 2+}(\chi)\right]\nonumber\\
&&+\frac{1}{\sqrt{2}}
       \Bigg[D^{(2)}_{A 1+}(\chi)  \frac{\xi_1}{\phi_2 - \phi_3}
             +D^{(2)}_{A 1-}(\chi) \frac{\xi_2}{\phi_3 - \phi_1}
             +D^{(2)}_{A 2-}(\chi)\frac{\xi_3}{\phi_1 - \phi_2}
       \Bigg]~,
\end{eqnarray}
with the intrinsic spin angular momentum densities
$\xi_i({\mathbf{x}})\equiv -R^{T}_{ij}(\chi({\mathbf{x}})) {\cal S}_j^{\rm spin}({\mathbf{x}})$,
\begin{eqnarray}
\quad\quad [{\cal S}_i^{\rm spin}({\mathbf{x}}),\xi_j({\mathbf{y}})]=0~,
\quad\quad [\xi_i({\mathbf{x}}),\xi_j({\mathbf{y}})]
    =-i\epsilon_{ijk}\delta^3({\mathbf{x}}-{\mathbf{y}})\xi_k({\mathbf{x}})~.
\nonumber
\end{eqnarray}
The spin vectors ${\cal S}_k^{\rm spin}$, finally, can be written as
\begin{eqnarray}
\label{new-spin1}
 {\cal S}_k^{\rm spin} &=&
       D^{(1)}_{k 1-}(\chi)  \xi_1
             +D^{(1)}_{k 1+}(\chi) \xi_2
             +D^{(1)}_{k 0}(\chi)\xi_3~.
\nonumber
\end{eqnarray}
Hence, in terms of the principal-axes variables, the part $H_0$ of the physical Hamiltonian,
containing no spatial derivatives, reads
\begin{eqnarray}
\label{H0pys}
H_0&=&\int d^3{\mathbf{x}}\ {1\over 2}\sum^{\rm
cyclic}_{i,j,k} \Bigg[ \pi_i^2 -{2i\delta({\mathbf{0}})\over \phi_j^2-
\phi_k^2}\!\left(\phi_j\pi_j-\phi_k \pi_k\right)
+\xi_i^2 {\phi_j^2+\phi_k^2\over (\phi_j^2-\phi_k^2)^2}
+\!  g^2 \phi_j^2 \phi_k^2\Bigg]~.
\end{eqnarray}
The matrix elements of a physical operator $O$ are given as
\begin{equation}
\langle\Psi'|O|\Psi\rangle \propto
 \prod_{\mathbf{x}}\int\!\!
d\alpha({\mathbf{x}}) \sin\beta d\beta({\mathbf{x}})
 d\gamma({\mathbf{x}})
% \!\!\!\!\!\!\!\!\!\!\!\!\!\!\!\!\!\!\!\!
\int\limits_{0<\phi_1({\mathbf{x}})<\phi_2({\mathbf{x}})<\phi_3({\mathbf{x}})}
%\!\!\!\!\!\!\!\!\!\!\!\!\!\!\!\!\!\!\!\!
\Big[\prod^{\rm cyclic}_{i,j,k}
d\phi_i({\mathbf{x}})
\! \left(\phi_j^2({\mathbf{x}})- \phi_k^2({\mathbf{x}})\right)\Big]
\Psi'^* O\Psi~.
\label{measure}
\end{equation}

\subsection{First and second order interaction terms}

The interaction parts of first and second order in the number of spatial derivatives, needed in this work,
can be written in the general form ($\Delta \equiv  \partial_x^2+\partial_y^2+\partial_z^2$)
\begin{eqnarray}
\label{partial}
V^{(\partial)}_\alpha &\equiv & C^{1 k}_{S_1 M_1\ S_2 M_2}\int d^3x\
\widetilde{Y}^{(S_1)}_{\alpha M_1}[\phi]\ i\partial_k Y^{(S_2)}_{\alpha M_2}[\phi]~,
\\
\label{Delta}
V^{(\Delta)}_\beta &\equiv & -\int d^3x\ \widetilde{X}^{(S)}_{\beta M}[\phi]
\Delta {X}^{(S)}_{\beta M}[\phi]
=V^{(0\Delta 0)}_\beta+V^{(1\Delta 1)}_\beta+V^{(2\Delta 2)}_\beta + ...~.
\end{eqnarray}
%\subsubsection{Magnetic terms}
In particular, the first order magnetic part reads
\begin{eqnarray}
\label{Vmagn1}
V^{(\partial)}_{\rm magn}\!\!\!\!&=&\!\!\!\!\
g\sqrt{\frac{5}{2}}\ C^{1 k}_{2 A\  2 B}\int d^3{\mathbf{x}}
   \left(\phi_1\phi_2\right)^{\!\!\!(2)}_{A} i\partial_k \left(\phi_3\right)^{\!\!\!(2)}_{B}~,
\end{eqnarray}
and the second order magnetic part is
\begin{eqnarray}
\label{Vmagn2}
\!\!\!\!\!\!\!\!\!\!\!\! V^{(\Delta)}_{\rm magn}\!\!\!\! & =&\!\!\!\!\! -\frac{1}{3}\! \int\!\! d^3{\mathbf{x}}
          \left[ \left(\phi_3\right)^{\!\!\!(0)}\!\! \Delta\!\left(\phi_3\right)^{\!\!\!(0)}
         \!\! + \!\left(\phi_3\right)^{\!\!\!(2)}_{A}\!\Delta\! \left(\phi_3\right)^{\!\!\!(2)}_{A}\right]
=V^{(0\Delta 0)}_{\rm magn}+V^{(2\Delta 2)}_{\rm magn}~.
\end{eqnarray}
From the second order term (\ref{Vmagn2}), we shall need in this work only the part $V^{(0\Delta 0)}_{\rm magn}$
and the expression
\begin{eqnarray}
 \widetilde{X}^{(S)}_{M} X^{(S)}_{M}\Big|_{\rm magn}
\!\!\!\! & =&\!\!\!\! \frac{1}{3}
          \left[ \left(\phi_3\right)^{\!\!\!(0)} \left(\phi_3\right)^{\!\!\!(0)}
          + \left(\phi_3\right)^{\!\!\!(2)}_{A} \left(\phi_3\right)^{\!\!\!(2)}_{A}\right]
=\frac{1}{\sqrt{3}}  \left(\phi_3^2\right)^{\!\!\!(0)}~.
\end{eqnarray}

%\subsubsection{Electric terms}

The first order electric term consists of transitions from spin-0 and spin-2 to spin-1
fields and therefore does not contribute.
Of the second order electric term I shall here only need ($\widehat{\phi}_i \equiv \phi_j+\phi_k~,
\ i,j,k\  {\rm cyclic})$
\begin{eqnarray}
V^{(0\Delta 0)}_{\rm elec}\!\!\!\! &=&\!\!\!\!
{1 \over 3 g^2}\int d^3{\mathbf{x}}
\Bigg[
\left({1\over \widehat{\phi}_3^2}\right)^{\!\!\!(0)}\!\!\!\!\Delta\left(T_3\right)^{\!\!(0)}
+{1 \over 4}\left({1\over \widehat{\phi}_3}\right)^{\!\!\!(0)}\!\!\!\!\Delta
\left({1\over \widehat{\phi}_1^3}\xi_1^2+{1\over \widehat{\phi}_2^3}\xi_2^2\right)^{\!\!\!(0)}
\nonumber\\
&&\quad\quad\quad\quad
-\left({1\over \widehat{\phi}_3^2}\ \widetilde{\widetilde{\pi}}_3\right)^{\!\!\!(0)}\!\!\!\!
\Delta \left(\widetilde{\widetilde{\pi}}_3\right)^{\!\!\!(0)}
-\delta({\bf 0})^2\left({1\over \widehat{\phi}_3^2}
\left({1\over \widehat{\phi}_1}+{1\over \widehat{\phi}_2}\right)\right)^{\!\!\!(0)}\!\!\!\!
\Delta\!\!\left({1\over \widehat{\phi}_3}\right)^{\!\!\!(0)}
\Bigg]~,
\end{eqnarray}
with the functions $(i,j,k\  {\rm cyclic})$
\begin{eqnarray}
T_i \!\!\!\!&=&\!\!\!\!{1 \over 2}\pi_i^2
+i\delta({\bf 0})\phi_i\left({1\over \phi_j^2-\phi_i^2}+{1\over \phi_k^2-\phi_i^2}\right)\pi_i
+{1 \over 4}{\phi_k^2+\phi_i^2\over \left(\phi_k^2-\phi_i^2\right)^2}\xi_j^2
+{1 \over 4}{\phi_j^2+\phi_i^2\over \left(\phi_j^2-\phi_i^2\right)^2}\xi_k^2~,
\nonumber\\
\widetilde{\widetilde{\pi}}_i \!\!\!\!&=&\!\!\!\!
\pi_i+i\delta({\bf 0})\phi_i\left({1\over \phi_j^2-\phi_i^2}+{1\over \phi_k^2-\phi_i^2}\right)~,
\nonumber
\end{eqnarray}
and
\begin{eqnarray}
\!\!\!\!\!\!\!\!  \widetilde{X}^{(S)}_{M} X^{(S)}_{M}\Big|_{\rm elec}
\!\!\!\!\!\!\!\!  & =&\!\!\!\!
\frac{1}{\sqrt{3}\ g^2}\Bigg[
    \left({1\over \widehat{\phi}_3^2}T_3\right)^{\!\!\!(0)}\!\!\!\!\!\!
+{1 \over 4}\left({1\over \widehat{\phi}_3}
\left({1\over \widehat{\phi}_1^3}\xi_1^2+{1\over \widehat{\phi}_2^3}\xi_2^2\right)\right)^{\!\!\!(0)}\!\!\!\!
-{1 \over 2}{1\over \widehat{\phi}_1\widehat{\phi}_2\widehat{\phi}_3}
\left({1\over \widehat{\phi}_3}\right)^{\!\!\!(0)}\!\!\!\!({\xi}_1^2+{\xi}_2^2+{\xi}_3^2)
\nonumber\\
&&\quad\quad\quad\quad
+\frac{\delta({\bf 0})^2}{8}
          \Bigg(
          21\left({1\over \widehat{\phi}_1^2\widehat{\phi}_2^2}\right)^{\!\!\!(0)}\!\!\!\!\!\!
          +11\left({1\over \widehat{\phi}_3^3}
          \left({1\over\widehat{\phi}_1}+{1\over\widehat{\phi}_2}\right)\right)^{\!\!\!(0)}\!\!\!\!\!\!
          +3{1\over\widehat{\phi}_1\widehat{\phi}_2\widehat{\phi}_3}\left({1\over \widehat{\phi}_3}
          \right)^{\!\!\!(0)}\!\!\Bigg)\Bigg]~.
\end{eqnarray}

%\subsubsection{Measure terms}

The first order measure part, containing spin-2 and spin-3 fields\footnote{
%------------------------------------------------------------------------------------
Using the notation
$\left(\phi_3\cdot\psi_3\right)^{(3)}_M:=
\left[
\left(\phi_2\psi_3-\phi_3\psi_2\right)
+\left(\phi_3\psi_1-\phi_1\psi_3\right)+\left(\phi_1\psi_2-\phi_2\psi_1\right)
\right]\left(D^{(3)}_{M\ 2}(\chi)-D^{(3)}_{M\ -2}(\chi)\right)/(2\sqrt{3})
$
}
%------------------------------------------------------------------------------------
, reads
\begin{eqnarray}
V^{(\partial)}_{\rm meas}\!\!\!\!&=&\!\!\!\!\frac{\delta(\mathbf{0})}{24 g}
\sqrt{\frac{5}{2}}\ C^{1 k}_{2 A\  2 B}
\!\!\int\!\! d^3{\mathbf{x}}   \Bigg[\!\!
\left[
   6 \left({1\over \widehat{\phi}_1\widehat{\phi}_2}\right)^{\!\!\!(2)}_{\!\!\! A}\!\!\!\!
   +3 \left({\widehat{\phi}_3\over \widehat{\phi}_1\widehat{\phi}_2}
\left({1\over \widehat{\phi}_1}+{1\over \widehat{\phi}_2}\right)\!\!\right)^{\!\!\!(2)}_{\!\!\! A}\!\!\!\!
-3 \left({1\over \widehat{\phi}_3}
\left({\widehat{\phi}_1\over \widehat{\phi}_2^2}
   +{\widehat{\phi}_2\over \widehat{\phi}_1^2}\right)\!\!\right)^{\!\!\!(2)}_{\!\!\! A}
\right]
   i\partial_k \left({1\over \widehat{\phi}_3}\right)^{\!\!\!(2)}_{B}
\nonumber\\
&&\quad\quad\quad\quad\quad\quad\quad\quad
+\left[
 3 \left({\widehat{\phi}_1\over \widehat{\phi}_2^2}
   +{\widehat{\phi}_2\over \widehat{\phi}_1^2}\right)^{\!\!\!(2)}_{\!\!\! A}\!\!\!\!
+\left({1\over \widehat{\phi}_3^2}
\left({\widehat{\phi}_1^2\over \widehat{\phi}_2}
   +{\widehat{\phi}_2^2\over \widehat{\phi}_1}\right)\!\!\right)^{\!\!\!(2)}_{\!\!\! A}\!\!\!\!
-\left({1\over \widehat{\phi}_3}
\left({\widehat{\phi}_1^2\over \widehat{\phi}_2^2}
   +{\widehat{\phi}_2^2\over \widehat{\phi}_1^2}\right)\!\!\right)^{\!\!\!(2)}_{\!\!\! A}
\right]
         i\partial_k
         \left({1\over \widehat{\phi}_1\widehat{\phi}_2}\right)^{\!\!\!(2)}_{B}
         \Bigg]
\nonumber\\
\!\!\!\!&&\!\!\!\!
+\frac{\delta(\mathbf{0})}{8 g}
\sqrt{\frac{7}{3}}\ C^{1 k}_{3 M\  2 B}\!\!
\int\!\! d^3{\mathbf{x}}\left[
3\left({1\over \widehat{\phi}_3^2}\cdot\widehat{\phi}_3\right)^{(3)}_M +
  \left({1\over \widehat{\phi}_3}
\left({1\over \widehat{\phi}_1^2}
   +{1\over \widehat{\phi}_2^2}\right)\cdot\widehat{\phi}_3^2\right)^{(3)}_M
\right]
     i\partial_k \left({1\over \widehat{\phi}_1\widehat{\phi}_2}\right)^{\!\!\!(2)}_{B}~,
\end{eqnarray}
Of the second order measure term, which is very complicated, I shall need here only
\begin{eqnarray}
V^{(0\Delta 0)}_{\rm meas}\!\!\!\! &=&\!\!\!\!
{\delta({\bf 0})^2 \over 48\ g^2}\!\!\int\!\! d^3{\mathbf{x}}
\Bigg[
\left[
-9\left({1\over \widehat{\phi}_3^2}\right)^{\!\!\!(0)}\!\!\!\!
            +8\left({\widehat{\phi}_3\over \widehat{\phi}_1\widehat{\phi}_2}
                  \left({1\over \widehat{\phi}_1}+{1\over \widehat{\phi}_2}\right)\right)^{\!\!\!(0)}\!\!\!\!
-\left({\widehat{\phi}_3^2\over \widehat{\phi}_1^2\widehat{\phi}_2^2}\right)^{\!\!\!(0)}
\right]
\Delta\ \left({1\over \widehat{\phi}_1 \widehat{\phi}_2}\right)^{\!\!\!(0)}
\nonumber\\
&&\quad\quad\quad
+{1 \over 5}\!\!
\left[\!
 14\!\left({1\over \widehat{\phi}_3^2}\right)^{\!\!\!(0)}\!\!\!\!\!
           +\! 42\!\left({1\over \widehat{\phi}_1 \widehat{\phi}_2}\right)^{\!\!\!(0)}\!\!\!\!\!
            -\! 5\!\left({\widehat{\phi}_3\over \widehat{\phi}_1\widehat{\phi}_2}
   \left({1\over \widehat{\phi}_1}+{1\over \widehat{\phi}_2}\right)\!\!\right)^{\!\!\!(0)}\!\!\!\!\!
+\! 4\!\left({\widehat{\phi}_3^2\over \widehat{\phi}_1^2\widehat{\phi}_2^2}\right)^{\!\!\!\!(0)}\!\!
\right]\!
\Delta\!\left[\!\!\left({1\over \widehat{\phi}_3^2}\right)^{\!\!\!(0)}\!\!\!\!
                               -\left({1\over \widehat{\phi}_1 \widehat{\phi}_2}\right)^{\!\!\!(0)}\!\!\right]
\nonumber\\
&&\quad\quad\quad\quad
-4\left[2\left({1\over \widehat{\phi}_3^3}\right)^{\!\!\!(0)}\!\!\!\!\!
+\left({1\over \widehat{\phi}_1\widehat{\phi}_2\widehat{\phi}_3}\right)^{\!\!\!(0)}\!\!
\right]\!
\Delta\!\left({1\over \widehat{\phi}_3}\right)^{\!\!\!(0)}
+{2 \over 5}
\Bigg[
2\left({1\over \widehat{\phi}_3^3}\right)^{\!\!\!(0)}\!\!\!\!
-\left({\widehat{\phi}_3\over \widehat{\phi}_1^2\widehat{\phi}_2^2}\right)^{\!\!\!(0)}\!\!\!\!
-2\left({1\over \widehat{\phi}_1\widehat{\phi}_2\widehat{\phi}_3}\right)^{\!\!\!(0)}\!\!\!\!
\nonumber\\
&&\quad\quad\quad\quad\quad\quad\quad\quad
-\left({1\over \widehat{\phi}_1\widehat{\phi}_2}
\left({1\over \widehat{\phi}_1}+{1\over \widehat{\phi}_2}\right)\!\!\right)^{\!\!\!(0)}\!\!\!\!
-\left({\widehat{\phi}_3\over \widehat{\phi}_1\widehat{\phi}_2}
\left({1\over \widehat{\phi}_1^2}+{1\over \widehat{\phi}_2^2}\right)\!\!\right)^{\!\!\!(0)}
\Bigg]
\Delta
\left[\left({\widehat{\phi}_3\over \widehat{\phi}_1\widehat{\phi}_2}\right)^{\!\!\!(0)}\!\!\!\!
      -\left({1\over \widehat{\phi}_3}\right)^{\!\!\!(0)}\right]
\nonumber\\
&&\quad\quad\quad\quad
+\left({1\over \widehat{\phi}_1\widehat{\phi}_2\widehat{\phi}_3}\right)^{\!\!\!(0)}\!\!\!\!\Delta
\left[
\left({\widehat{\phi}_3^2\over \widehat{\phi}_1\widehat{\phi}_2}
\left({1\over \widehat{\phi}_1}+{1\over \widehat{\phi}_2}\right)\!\!\right)^{\!\!\!(0)}\!\!\!\!
- \left(\widehat{\phi}_3
\left({1\over \widehat{\phi}_1^2}+{1\over \widehat{\phi}_2^2}\right)\!\!\right)^{\!\!\!(0)}
\right]
\Bigg]~,
\end{eqnarray}
and
\begin{eqnarray}
  \widetilde{X}^{(S)}_{M} X^{(S)}_{M}\Big|_{\rm meas}\!\!\!\! & =&\!\!\!\!
          -\frac{\delta({\bf 0})^2}{8\sqrt{3}\ g^2}\int d^3{\mathbf{x}}
          \Bigg[3\left({1\over \widehat{\phi}_1^2\widehat{\phi}_2^2}\right)^{\!\!\!(0)}
          +6\left({1\over \widehat{\phi}_3^3}
          \left({1\over\widehat{\phi}_1}+{1\over\widehat{\phi}_2}\right)\right)^{\!\!\!(0)}
          \Bigg]~.
\end{eqnarray}

\section{Coarse graining and strong coupling expansion in $\lambda=g^{-2/3}$}

I now set an ultraviolet cutoff $a$ by introducing an infinite spatial lattice of granulas
$G({\mathbf{n}},a)$, here cubes of length $a$,
situated at sites ${\mathbf{x}}=a {\mathbf{n}}$  $({\mathbf{n}}=(n_1,n_2,n_3)\in Z^3)$, and
considering the averaged variables
\begin{equation}
\label{average}
\phi({\mathbf{n}}) :=  \frac{1}{a^3}\int_{G({\mathbf{n}},a)} d{\mathbf{x}}\ \phi({\mathbf{x}})
\nonumber
\end{equation}
(where in particular $\delta({\bf 0})\rightarrow 1/a^3$),
and the discretized first and second spatial derivatives (s=1,2,3),
\begin{eqnarray}
\label{1stdisc}
\partial_s\phi({\mathbf{n}})& := &\lim_{N\rightarrow\infty}\sum_{n=1}^N w_N(n)
\frac{1}{2na}\left(\phi({\mathbf{n}}+n {\mathbf{e}}_s)
     -\phi({\mathbf{n}}-n {\mathbf{e}}_s)\right)\\
%------------------------------------------------------------------------------
\partial_s^2\phi({\mathbf{n}}) & := &\lim_{N\rightarrow\infty}\sum_{n=1}^N w_N(n)
\frac{1}{(n a)^2 }
\Big(\phi({\mathbf{n}}+n {\mathbf{e}}_s)
     +\phi({\mathbf{n}}-n {\mathbf{e}}_s)
     -2\phi({\mathbf{n}})\Big)
\label{2nddisc}
\end{eqnarray}
with the unit lattice vectors ${\mathbf{e}}_1=(1,0,0),{\mathbf{e}}_2=(0,1,0),{\mathbf{e}}_3=(0,0,1)$
and the distribution
\begin{equation}
\label{distribution}
w_N(n):=2\frac{(-1)^{n+1}(N!)^2}{(N-n)!(N+n)!}~,\quad 1\leq n\leq N~,\quad\sum_{n=1}^N w_N(n)= 1~.
\end{equation}
%-----------------------------------------------------------------------------------------
 The values of $\partial_s\phi ({\mathbf{n}})$ and $\partial_s^2\phi ({\mathbf{n}})$ in (\ref{1stdisc})
and (\ref{2nddisc}) for a given site ${\mathbf{n}}$ and direction, say $s=1$, are chosen
to coincide with the first and second derivative,
$I'_{2N}(a n_1)|_{n_2,n_3}$ and $I''_{2N}(a n_1)|_{n_2,n_3}$
\footnote{
%----------------------------------------------------------------------------------
Differentiating the Lagrange interpolation
polynomials $I_{2N}(x)$ with given values
$y_n$ at the equidistant points $x_n=x_0+ n a$, ($n=-N,-N+1,..,N-1,N$),
once/twice at the central point $x_0$, one obtains: $I'_{2N}(x_0)=\sum_{n=1}^N w_N(n)(y_{n}-y_{-n})/(2na)$ and
$I''_{2N}(x_0)=\sum_{n=1}^N w_N(n)(y_{n}+y_{-n}-2y_0)/(na)^2$ with the distribution (\ref{distribution}).
For N=1, in particular, one has $I'_{2}(x_0)=(y_{1}-y_{-1})/(2a)$ and
$I''_{2}(x_0)=(y_{1}+y_{-1}-2y_0)/a^2$
%----------------------------------------------------------------------------------
}
,respectively,
of the interpolation polynomial $I_{2N}(x_1)|_{n_2,n_3}$ in the $x_1$ coordinate,
which is uniquely determined by the series of values $\phi(n_1+n,n_2,n_3)$ $(n=-N,..,N)$ obtained via
the averaging (\ref{average}),
and then taking the limit $N \rightarrow \infty$.
%The interpolating function therefore plays the role of a smoothened version of the original field
%$\phi({\mathbf{x}})$, and the discretised values $\phi({\mathbf{n}})$,
% $\partial_s\phi ({\mathbf{n}})$, and $\partial_s^2\phi ({\mathbf{n}})$ in (\ref{average})- (\ref{2nddisc}),
% are just the values and the partial derivatives
%of this smoothened field at lattice site ${\mathbf{n}}$.
Note, that the ($N=1$) choice,
$\partial_s\phi({\mathbf{n}})|_{N=1} =\left(\phi({\mathbf{n}}+ {\mathbf{e}}_s)
     -\phi({\mathbf{n}}- {\mathbf{e}}_s)\right)/(2a)$ and
$\partial_s^2\phi({\mathbf{n}})|_{N=1} =\left(\phi({\mathbf{n}}+ {\mathbf{e}}_s)
     +\phi({\mathbf{n}}- {\mathbf{e}}_s)
     -2\phi({\mathbf{n}})\right)/a^2$,
which includes only the nearest neighbors ${\mathbf{n}}\pm {\mathbf{e}}_s$,
would lead to the same results as (\ref{1stdisc}) and (\ref{2nddisc})
for the soft components of the original field $\phi({\mathbf{x}})$,
which vary only slightly over several lattice sites, but lead to values falling off
faster than (\ref{1stdisc}) and (\ref{2nddisc})
for higher momentum components approaching $\pi/a$.

Applying furthermore the rescaling transformation (afterwards again dropping the primes)
\begin{equation}
\phi_i = \frac{g^{-1/3}}{a} \phi_i^{\prime}~, \quad\quad
\pi_i  = \frac{g^{1/3}}{a^2}\pi_i^{\prime}~,  \quad\quad
\xi_i  = \frac{1}{a^3}      \xi_i^{\prime}~,
\end{equation}
I obtain the expansion of the Hamiltonian in $\lambda=g^{-2/3}$
\begin{equation}
H =\frac{g^{2/3}}{a}\left[{\cal H}_0+\lambda \sum_\alpha {\cal V}^{(\partial)}_\alpha
                             +\lambda^2 \left(\sum_\beta {\cal V}^{(\Delta)}_\beta
                             +\sum_\gamma{\cal V}^{(\partial\partial\neq\Delta)}_\gamma\right)
                             + {\mathcal{O}}(\lambda^3)\right]~,
\end{equation}
with the "free" Hamiltonian
\begin{eqnarray}
\label{calH0}
{\cal H}_0 &=&
\sum_{\mathbf{n}}\Bigg[{1\over 2}\sum^{\rm cyclic}_{ijk}
\Big[ \pi_i^2({\mathbf{n}})
-{2i\over \phi_j^2({\mathbf{n}})-\phi_k^2({\mathbf{n}})}
\left(\phi_j({\mathbf{n}})\pi_j({\mathbf{n}})
-\phi_k({\mathbf{n}}) \pi_k({\mathbf{n}})\right)
\nonumber\\
&&\quad\quad\quad\quad\quad\quad\quad\quad
+ \xi^2_i({\mathbf{n}}) {\phi_j^2({\mathbf{n}})+\phi_k^2({\mathbf{n}})
\over (\phi_j^2({\mathbf{n}})-\phi_k^2({\mathbf{n}}))^2}
 +\phi_j^2({\mathbf{n}}) \phi_k^2({\mathbf{n}})
\Big]\Bigg]=\sum_{\mathbf{n}}{\cal H}^{QM}_0({\mathbf{n}})~,
\end{eqnarray}
which is the sum of the Hamiltonians of $SU(2)$-Yang-Mills quantum mechanics of constant
fields in each box, and the interaction parts, relating different boxes,
\begin{eqnarray}
\label{calVp}
\!\!\!\!\!\!\!\!\!
{\cal V}^{(\partial)}_\alpha \!\!\!\!\!\!& =&\!\!\!\!\!\!
           \lim_{N\rightarrow\infty}\sum_{n=1}^N w_N(n)\Bigg[
\frac{i}{2n} C^{1 s}_{S_1 M_1\ S_2 M_2}\sum_{\mathbf{n}}
\widetilde{\cal Y}^{(S_1)}_{\alpha M_1}[\phi({\mathbf{n}})]
\left({\cal Y}^{(S_2)}_{\alpha M_2}[\phi({\mathbf{n}}+n{\mathbf{e}}_s)]
     -{\cal Y}^{(S_2)}_{\alpha M_2}[\phi({\mathbf{n}}-n {\mathbf{e}}_s)]\right)\!\!\Bigg]
\\
\label{calVpp}
\!\!\!\!\!\!\!\!\!
{\cal V}^{(\Delta)}_\beta \!\!\!\!\!\!&=&\!\!\!\!
        -\!\!\!\lim_{N\rightarrow\infty}\!\sum_{n=1}^N\! w_{N}(n)\!\Bigg[\!\frac{1}{n^2}\!
\sum_{{\mathbf{n}},s}\widetilde{\cal X}^{(S)}_{\beta M}[\phi({\mathbf{n}})]
\Big({\cal X}^{(S)}_{\beta M}[\phi({\mathbf{n}}+n{\mathbf{e}}_s)]\!
     +\! {\cal X}^{(S)}_{\beta M}[\phi({\mathbf{n}}-n{\mathbf{e}}_s)]\!
     -2\!{\cal X}^{(S)}_{\beta M}[\phi({\mathbf{n}})]\Big)\!\!\Bigg]
\end{eqnarray}
with the dimensionless and coupling constant independent terms ${\cal X},{\cal Y}$,
obtained from the $X,Y$ in (\ref{partial}) and (\ref{Delta})
by putting ${\cal X}[\phi]:=X[\phi]|_{a=1,g=1,\delta(0)=1}$ and
${\cal Y}[\phi]:=Y[\phi]|_{a=1,g=1,\delta(0)=1}$.

The expansion of the Hamiltonian in terms of the number of spatial derivatives
is therefore equivalent to a strong coupling expansion in $\lambda=g^{-2/3}$.
It is the analogon
of the weak coupling expansion in $g^{2/3}$ for small boxes
by L\"uscher and M\"unster\cite{Luescher},\cite{Luescher and Muenster}\footnote{\label{smallbox}
%-------------------------------------------------------------------------------
Integrating out all higher modes in a small box of size a, a weak coupling expansion for energies
of the constant fields, $E = \frac{1}{a}\sum_{k=0}^\infty \epsilon_k \bar{\lambda}^k~,
\  \bar{\lambda}\equiv[\bar{g}(\Lambda_{MS}a)]^{2/3}$ is obtained, with the standard running coupling
constant in the MS scheme.
%-------------------------------------------------------------------------------
},
and supplies a useful alternative to strong coupling expansions based on the
Wilson-loop gauge invariant variables, which had been carried out by Kogut, Sinclair, and Susskind
\cite{Kogut} for a 3-dimensional spatial lattice in the Hamiltonian formalism,
yielding an expansion in $1/g^4$, and by M\"unster \cite{Muenster} for a
4-dimensional space-time lattice.

The low energy spectrum and eigenstates of ${\cal H}_0^{QM}$  at each site ${\mathbf{n}}$
appearing in (\ref{calH0}),
\begin{equation}
{\cal H}_0^{QM}({\mathbf{n}})  |\Phi_{i,M}^{(S)}\rangle_{\mathbf{n}} =
\epsilon^{(S)}_i ({\mathbf{n}}) |\Phi_{i,M}^{(S)}\rangle_{\mathbf{n}}~,
\end{equation}
characterised by the quantum numbers of spin $S,M$,
have been obtained in
\cite{Luescher and Muenster},\cite{Koller and van Baal},\cite{pavel}
with high accuracy.
It is important to note (see \cite{pavel} for details), that at strong coupling,
due to the positivity of the range (\ref{range}),
all states should satisfy either the $(+)$ b.c. $\partial_{\phi_1}\!\!\Phi(\phi)|_{\phi_1=0}=0$,
or the  $(-)$ b.c. $\Phi(\phi)|_{\phi_1=0}=0$,
in accordance with (\ref{range}) and the
invariance  of the Hamiltonian ${\cal H}_0$ under parity transformation $\phi\rightarrow -\phi$.
The spectrum is purely discrete in both cases and the lowest energies are
\begin{equation}
\epsilon_{0}^{+}=4.1167~,\quad
\epsilon_{0}^{-}=8.7867~.
\end{equation}
The energies (relative to $\epsilon_{0}$)
\begin{equation}
\mu_{i}^{(S)+}:=\epsilon_{i}^{(S)+}-\epsilon_{0}^{+}~,\quad
\mu_{i}^{(S)-}:=\epsilon_{i}^{(S)-}-\epsilon_{0}^{-}~,
\end{equation}
of the lowest states for spin-0,2,3 and 4 for $(+)$ and $(-)$ b.c. are summarized in Table 1a
and 1b.
Spin-1 states are absent for both cases.
The underlined values correspond to stable excitations below
threshold
\begin{equation}
\label{threshold}
\mu_{\rm th}^{+}=3.796\ \ (=2 \mu_{1}^{(2)+})~,\quad
\mu_{\rm th}^{-}=5.089\ \ (=2 \mu_{1}^{(2)-})~,
\end{equation}
for decay into two spin-2 excitations $\mu_{1}^{(2)}$ (lightest in the spectrum).
\newline
\newline
$\begin{array}{|c||c|c|c|c|}
\mu_{i}^{(S)+} & S=0 & S=2 & S=3 & S=4 \\  \hline\hline
i=1 & \underline{2.270}  & \underline{1.898} & 8.009 & \underline{3.61}\\ \hline
i=2 & 3.857 & \underline{3.704} & 10.815 & 5.23\\ \hline
i=3 & 5.09 & 5.22 & 13.1 & 6.9\\ \hline
\end{array}$\quad
$\begin{array}{|c||c|c|c|c|}
\mu_{i}^{(S)-} & S=0 & S=2 & S=3 & S=4 \\  \hline\hline
i=1 & \underline{3.268} & \underline{2.545} & 9.250 & \underline{4.93}\\ \hline
i=2 & 5.233 & 5.212 & 12.78 & 7.37\\ \hline
i=3 & 6.803 & 6.612 & 15.38 & 9.6\\ \hline
\end{array}$
\newline
\newline
Table 1a and 1b: Results for the first three excitation energies $\mu_{i}^{(S)}$
for(+) and (-) b.c. The underlined values correspond to stable excitations below
threshold (\ref{threshold}). The numerical errors (estimated from the deviation from the virial
theorem, see \cite{pavel}) are smaller than the last digit in the numbers given.

\section{Perturbation theory in $\lambda=g^{-2/3}$}

\subsection{ Free many-glueball states}

The eigenstates of the free Hamiltonian
\begin{eqnarray}
H_0=\frac{g^{2/3}}{a}\sum_{\mathbf{n}}{\cal H}^{QM}_0({\mathbf{n}})
 \nonumber
\end{eqnarray}
are free many-glueball states (completely decoupled granulas).
The free glueball vacuum  is
\begin{eqnarray}
 |0\rangle \equiv  \bigotimes_{\mathbf{n}} |\Phi_0\rangle_{\mathbf{n}} \
 \rightarrow\  E_{\rm vac}^{\rm free}={\cal N} \epsilon_0\frac{g^{2/3}}{a}
 \nonumber
\end{eqnarray}
(${\cal N}$ total number of granulas) with all granulas in the lowest state of energy $\epsilon_0$.
The free one-glueball states, which in this work I choose to be momentum eigenstates, are
\begin{eqnarray}
 |S,M,i,{\mathbf{k}}\rangle &\equiv &  \sum_{\mathbf{n}} e^{i a {\mathbf{k}}.{\mathbf{n}}}
 \left[ |\Phi_{i,M}^{(S)}\rangle_{\mathbf{n}} \bigotimes_{{\mathbf{m}}\neq {\mathbf{n}}}
 |\Phi_0\rangle_{\mathbf{m}}\right]\
\rightarrow \
 E_i^{(S){\rm free}}(k)
=\mu_i^{(S)}\frac{g^{2/3}}{a} + E_{\rm vac}^{\rm free}~,
\nonumber
\end{eqnarray}
the free two-glueball states,
\begin{eqnarray}
|(S_1,M_1,i_1,{\mathbf{n}}_1),(S_2,M_2,i_2,{\mathbf{n}}_2)\rangle &\equiv &
 |\Phi_{i_1,M_1}^{(S_1)}\rangle_{{\mathbf{n}}_1}
 \otimes|\Phi_{i_2,M_2}^{(S_2)}\rangle_{{\mathbf{n}}_2}
\left[ \bigotimes_{{\mathbf{m}}\neq {\mathbf{n}}_1,{\mathbf{n}}_2}
 |\Phi_0\rangle_{\mathbf{m}}\right]\nonumber\\
\rightarrow \
 E_{i_1,i_2}^{(S_1,S_2){\rm free}}
&=&(\mu_{i_1}^{(S_1)}+\mu_{i_2}^{(S_2)})\frac{g^{2/3}}{a}
+ E_{\rm vac}^{\rm free}~,\nonumber
\end{eqnarray}
and so on.
Matrix elements between these free glueball states are calculated using the measure (\ref{measure}).

\subsection{Interacting glueball vacuum}

The energy of the interacting glueball vacuum up to $\lambda^2$
\begin{eqnarray}
\!\!\!\!\!\!\!\!\!\!\!\!\!\!\!\!\!\!\!\!\!\!\!\!\!
E_{\rm vac}\!\!\!\!\!&=&\!\!\!\!\! {\cal N}\frac{g^{2/3}}{a}
\Bigg[\epsilon_0+
\lambda^2
\sum_\beta \langle 0 | {\cal V}^{(\Delta)}_\beta|0\rangle
-\lambda^2\Bigg(\sum_{\alpha,\alpha^\prime}\sum_{\ i_1, i_2}
\frac{
\langle 0 | {\cal V}^{(\partial)}_{\alpha,2-2}|2_{i_1} 2_{i_2}\rangle\langle 2_{i_1} 2_{i_2}|
{\cal V}^{ (\partial)}_{\alpha^\prime,2-2}|0 \rangle}
{\mu_{i_1}^{(2)}+\mu_{i_2}^{(2)}}
\nonumber\\
&&\quad\quad\quad\quad\quad
+\sum_{i_1,i_2}
\frac{
\langle 0 | {\cal V}^{(\partial)}_{2-3}|2_{i_1} 3_{i_2}\rangle\langle 2_{i_1} 3_{i_2}|
{\cal V}^{\prime (\partial)}_{2-3}|0 \rangle}
{\mu_{i_1}^{(2)}+\mu_{i_2}^{(3)}}
\Bigg)\!\! +{\cal O}(\lambda^3)
\Bigg]
\equiv\frac{g^{2/3}}{a}\Bigg[\epsilon_0+c_0\lambda^2+{\cal O}(\lambda^3)\Bigg]
\end{eqnarray}
is obtained using first and second order perturbation theory.

%\subsubsection{First order perturbation theory}

For any ${\cal V}^{(\Delta)}_\beta$ of (\ref{calVpp}) I obtain,
using $\lim_{N\rightarrow\infty}\sum_{n=1}^N (w_N(n)/ n^2)=\zeta(2)=\pi^2/6$~,
\begin{eqnarray}
\label{vac1st}
c_0\Big|^{\rm 1st\ ord}_\beta &=&\langle 0 | {\cal V}^{(\Delta)}_\beta|0\rangle=
  \pi^2\Bigg[ \langle \Phi_0|
   \left(
   \widetilde{\cal X}^{(S)}_{\beta,M}
    {\cal X}^{(S)}_{\beta,M}\right)
                      |\Phi_0\rangle
           -\langle \Phi_0|\widetilde{\cal X}^{(0)}_\beta |\Phi_0\rangle
               \langle \Phi_0|{\cal X}^{(0)}_\beta |\Phi_0\rangle
\Bigg]~.
\end{eqnarray}
For example, for the magnetic potential ${\cal V}^{(\Delta)}_{\rm magn}$, corresponding to the
$V^{(\Delta)}_{\rm magn}$ in (\ref{Vmagn2}), Eq. (\ref{vac1st}) becomes
\begin{eqnarray}
c_0\Big|_{\rm magn}^{\rm 1st\ ord}\!\!\!\!&=&\!\!\!\! \frac{\pi^2}{3}\Big[
 \langle \Phi_0|\phi_1^2+\phi_2^2+\phi_3^2|\Phi_0\rangle
 -\frac{1}{3}\langle \Phi_0|\phi_1+\phi_2+\phi_3|\Phi_0\rangle^2\Big]
= 4.560 \ (3.514)\nonumber~.
\end{eqnarray}
Here (and in the following paragraphs) the number without brackets corresponds to the $(+)$ b.c. and the
number in brackets to the $(-)$ b.c.
The numerical errors are smaller than the last digit in the numbers given.
Together with the corresponding contributions $42.323\ (13.229)$ for the electric
${\cal V}^{(\Delta)}_{\rm elec}$ and $-16.408 \ (-1.782)$ for the measure terms
${\cal V}^{(\Delta)}_{\rm meas}$, I find the total first order
\begin{eqnarray}
c_0\Big|_{\rm tot}^{\rm 1st\ ord}\!\!\!\!&=&\! 30.474\ (14.962) \nonumber ~.
\end{eqnarray}

%\subsubsection{Second order perturbation theory}

Using ${\cal V}^{(\partial)}_\alpha$ in (\ref{calVp}) and
$\lim_{N\rightarrow\infty}\sum_{n=1}^N (w^2_N(n)/ n^2)=4\zeta(2)=2\pi^2/3$,
I obtain for the contribution due to the vacuum polarization
into a virtual pair of spin-2 particles ,
%%%%%%%%%%%%%%%%%%%%%%%%%%%%%%%%%%%%%%%%%%%%%%%%%%%%%%%%%%%%%%%%%%%%%%%%%%%%%%%%%%%%%%%%%%%%
\begin{eqnarray}
\label{vac2nd}
c_0\Bigg|^{\rm 2nd\ ord}_{2-2,\alpha,\alpha^{\prime}}\!\!\!\!\!\!\!\!\!\!\!\!\!\!
 &=&-\frac{\pi^2}{25}
\sum_{i_1,i_2}
\frac{   \langle \Phi_0||\widetilde{\mathbf{\cal Y}}^{(2)}_\alpha||\Phi_{i_1}^{(2)}\rangle
         \langle \Phi_0||{\mathbf{\cal Y}}^{(2)}_\alpha ||\Phi_{i_2}^{(2)}\rangle}
   {\mu_{i_1}^{(2)}+\mu_{i_2}^{(2)}}
 \Bigg[
      \langle \Phi_{i_1}^{(2)}||\widetilde{\mathbf{\cal Y}}^{(2)}_{\alpha^{\prime}}||\Phi_0\rangle
      \langle \Phi_{i_2}^{(2)}||{\mathbf{\cal Y}}^{(2)}_{\alpha^{\prime}}||\Phi_0\rangle
\nonumber\\
 &&\quad\quad\quad\quad\quad\quad\quad\quad\quad\quad\quad\quad\quad\quad\quad\quad\quad\quad
\quad\quad
+\langle \Phi_{i_2}^{(2)}||\widetilde{\mathbf{\cal Y}}^{(2)}_{\alpha^{\prime}}||\Phi_0\rangle
  \langle \Phi_{i_1}^{(2)}||{\mathbf{\cal Y}}^{(2)}_{\alpha^{\prime}} ||\Phi_0\rangle
\Bigg]~,
\end{eqnarray}
and similarly that due to the vacuum polarization into a spin-2 and a spin-3 particle,
\begin{eqnarray}
c_0\Bigg|^{\rm 2nd\ ord}_{2-3}\!\!\!\!\!\!\!\!\!\!\!\!\!\!
 &=&-\frac{\pi^2}{35}
\sum_{i_1,i_2}
\frac{   \langle \Phi_0||\widetilde{\mathbf{\cal Y}}^{(3)}||\Phi_{i_1}^{(3)}\rangle
         \langle \Phi_0||{\mathbf{\cal Y}}^{(2)} ||\Phi_{i_2}^{(2)}\rangle}
   {\mu_{i_1}^{(3)}+\mu_{i_2}^{(2)}}
      \langle \Phi_{i_1}^{(3)}||\widetilde{\mathbf{\cal Y}}^{\prime(3)}||\Phi_0\rangle
      \langle \Phi_{i_2}^{(2)}||{\mathbf{\cal Y}}^{\prime(2)}||\Phi_0\rangle~.
\nonumber
\end{eqnarray}
The leading contribution to (\ref{vac2nd}) comes from the
${\cal V}^{(\partial)}_{\rm magn}-{\cal V}^{(\partial)}_{\rm magn}$ vacuum polarization
(see $V^{(\partial)}_{\rm magn}$ in (\ref{Vmagn1}))
\begin{eqnarray}
c_0\Bigg|^{\rm 2nd\ ord}_{\rm magn-magn}\!\!\!\!\!\!\!\!\!\!\!\!\!\!
 &=&-\frac{\pi^2}{10}
\sum_{i_1,i_2}
\frac{   \langle \Phi_0||\left(\phi_1\phi_2\right)^{\!(2)}||\Phi_{i_1}^{(2)}\rangle
         \langle \Phi_0||\left(\phi_3\right)^{\!(2)} ||\Phi_{i_2}^{(2)}\rangle}
   {\mu_{i_1}^{(2)}+\mu_{i_2}^{(2)}}
 \Bigg[
      \langle \Phi_{i_1}^{(2)}||\left(\phi_1\phi_2\right)^{\!(2)}||\Phi_0\rangle
      \langle \Phi_{i_2}^{(2)}||\left(\phi_3\right)^{\!(2)}||\Phi_0\rangle
\nonumber\\
 &&\quad\quad\quad\quad\quad\quad\quad\quad\quad
+\langle \Phi_{i_2}^{(2)}||\left(\phi_1\phi_2\right)^{\!(2)}||\Phi_0\rangle
  \langle \Phi_{i_1}^{(2)}||\left(\phi_3\right)^{\!(2)}||\Phi_0\rangle
\Bigg]=-0.399\ (-0.341)~.
\nonumber
\end{eqnarray}
Together with the smaller contributions  $-0.1516 \ (-0.0186)$ from
${\cal V}^{(\partial)}_{\rm magn}-{\cal V}^{(\partial)}_{\rm meas}$, $-0.0295 \ (-0.0004)$ from
 ${\cal V}^{(\partial)}_{\rm meas}-{\cal V}^{(\partial)}_{\rm meas}$,
 and the negligibly small $-3.5 \times 10^{-5} \ (-8.6 \times 10^{-5})$ from
 ${\cal V}^{(\partial)}_{2-3}-{\cal V}^{(\partial)}_{2-3}$,
 I find the total second order
\begin{eqnarray}
c_0\Big|_{\rm tot}^{\rm 2nd\ ord}\!\!\!\!&=&\! -0.580\ (-0.360) \nonumber ~.
\end{eqnarray}
Hence 1st and 2nd order perturbation theory together give the result
\begin{eqnarray}
E_{\rm vac}^{+}={\cal N}\frac{g^{2/3}}{a}\Bigg[4.1167+29.894\lambda^2+{\cal O}(\lambda^3)\Bigg]~,\quad
E_{\rm vac}^{-}={\cal N}\frac{g^{2/3}}{a}\Bigg[8.7867+14.602\lambda^2+{\cal O}(\lambda^3)\Bigg]~,
\end{eqnarray}
for the energy of the interacting glueball vacuum up to $\lambda^2$,
for the $(+)$ and $(-)$ boundary conditions, respectively.
The results are summarized in Table 2.
\newline
\newline
$\begin{array}{|c||c|c|c|c|c|}
{\rm vacuum} & \epsilon_0 & c_0^{({\rm 1st})} & c_0^{({\rm 2nd})}
& c_0 & c_0/\epsilon_0 \\  \hline\hline
(+) & 4.1167 & 30.474 & -0.580 & 29.894 & 7.262  \\ \hline
(-) & 8.7867 & 14.962 & -0.360 & 14.602 & 1.662  \\ \hline
\end{array}$ \newline
\newline
Table 2: Results for the interacting glueball vacuum  for (+) and (-) b.c.
The numerical errors are smaller than the last digits in the numbers shown.

\subsection{Interacting Spin-0 glueballs}

Including interactions $V^{\Delta}$ and $V^{\partial}$ using 1st and 2nd order perturbation theory,
we obtain the following energy of the interacting spin-0 glueball up to $\lambda^2$,
\begin{eqnarray}
E_i^{(0)}(k)-E_{\rm vac}\!\!\!\!\!&=&\!\!\!\!\!\frac{g^{2/3}}{a}
\Bigg[\mu_i^{(0)}\!\! +\!\!
\lambda^2\!\sum_\beta
\langle 0 i k | {\cal V}^{(\Delta)}_\beta|0 i k\rangle\!
-\lambda^2\!\Bigg(\sum_{\alpha,\alpha^\prime}\!\!\!\!\!\!\sum_{\ \ \ \ i_1, i_2}\!\!\!\!\!\!
\frac{
\langle 0 i k | {\cal V}^{(\partial)}_{\alpha,2-2}|2_{i_1} 2_{i_2}\rangle\langle 2_{i_1} 2_{i_2}|
{\cal V}^{ (\partial)}_{\alpha^\prime,2-2}|0 i k \rangle}
{\mu_{i_1}^{(2)}+\mu_{i_2}^{(2)}-\mu_{i}^{(0)}}
\nonumber\\
&&
\quad\quad\quad\quad
+\sum_{i_1,i_2}
\frac{
\langle 0 i k | {\cal V}^{(\partial)}_{2-3}|2_{i_1} 3_{i_2}\rangle\langle 2_{i_1} 3_{i_2}|
{\cal V}^{\prime (\partial)}_{2-3}|0 i k \rangle}
{\mu_{i_1}^{(2)}+\mu_{i_2}^{(3)}-\mu_{i}^{(0)}}
\Bigg)+{\cal O}(\lambda^3)\Bigg]\nonumber\\
&\equiv &
 \frac{g^{2/3}}{a}\left[\mu_i^{(0)}
 + \lambda^2\left( c_i^{(0)}+  \widetilde{c}_i^{(0)} a^2 k^2 +{\cal O}((a^2 k^2)^2)\right)
 +{\cal O}(\lambda^3)\right]
\label{spin0sp}
\end{eqnarray}
All spin-0 glueball excitations
are unstable at tree-level, except for the lowest $\mu_1^{(0)}$,
which is below threshold (\ref{threshold}) for decay into two spin-2 glueballs .

%\subsubsection{First order perturbation theory}

For a potential term of the general form ${\cal V}^{(\Delta)}_\beta$ of (\ref{calVpp}),
I find in first order perturbation theory
\begin{eqnarray}
\label{spin01sta}
c_1^{(0)}\Big|^{\rm 1st\ ord}_\beta \!\!\!\!\!\!\!\!\!\!\!\!&=&\!\!\!\!
  \pi^2\Bigg[\langle \Phi_{1}^{(0)}|
  \left(
  \widetilde{\cal X}^{(S)}_{\beta,M}
   {\cal X}^{(S)}_{\beta,M}\right)
                        |\Phi_{1}^{(0)}\rangle
- \langle \Phi_0|
   \left(
   \widetilde{\cal X}^{(S)}_{\beta,M}
    {\cal X}^{(S)}_{\beta,M}\right)
                      |\Phi_0\rangle
\nonumber\\
&&\!\!\!\!\!\!\!\!
          -\left( \langle \Phi_1^{(0)}|\widetilde{\cal X}^{(0)}_\beta |\Phi_1^{(0)}\rangle
                 -\langle \Phi_0|\widetilde{\cal X}^{(0)}_\beta |\Phi_0\rangle \right)
                 \langle \Phi_0| {\cal X}^{(0)}_\beta|\Phi_0\rangle
           -\langle \Phi_0|\widetilde{\cal X}^{(0)}_\beta |\Phi_0\rangle
               \left( \langle \Phi_1^{(0)}|{\cal X}^{(0)}_\beta |\Phi_1^{(0)}\rangle
                 -\langle \Phi_0|{\cal X}^{(0)}_\beta |\Phi_0\rangle \right)
\nonumber\\
&&
           -\langle \Phi_{1}^{(0)}| \widetilde{\cal X}^{(0)}_\beta|\Phi_0\rangle
                \langle \Phi_0|{\cal X}^{(0)}_\beta |\Phi_{1}^{(0)}\rangle
                -\langle \Phi_0|\widetilde{\cal X}^{(0)}_\beta|\Phi_{1}^{(0)}\rangle
                \langle \Phi_{1}^{(0)}| {\cal X}^{(0)}_\beta|\Phi_0\rangle\Bigg]~,
\\
\label{spin01stb}
c_1^{(0)}\Big|^{\rm 1st\ ord}_\beta \!\!\!\!\!\!\!\!\!\!\!\!&=&\!\!\!\!
        \Bigg[\langle \Phi_{1}^{(0)}| \widetilde{\cal X}^{(0)}_\beta|\Phi_0\rangle
                \langle \Phi_0|{\cal X}^{(0)}_\beta |\Phi_{1}^{(0)}\rangle
                +\langle \Phi_0|\widetilde{\cal X}^{(0)}_\beta|\Phi_{1}^{(0)}\rangle
                \langle \Phi_{1}^{(0)}| {\cal X}^{(0)}_\beta|\Phi_0\rangle
\Bigg]~.
\end{eqnarray}
For example, for the magnetic potential ${\cal V}^{(\Delta)}_{\rm magn}$ corresponding to the
$V^{(\Delta)}_{\rm magn}$ in (\ref{Vmagn2}),
Equs. (\ref{spin01sta}) and (\ref{spin01stb}) give
\begin{eqnarray}
c_1^{(0)}\Big|_{\rm magn}^{\rm 1st\ ord}\!\!\!\!\!\!\!\!&=&\!\!\!\! \frac{\pi^2}{3}\Big[
 \left(\langle \Phi_1^{(0)}|\phi_1^2+\phi_2^2+\phi_3^2|\Phi_i^{(0)}\rangle
 -\langle \Phi_0|\phi_1^2+\phi_2^2+\phi_3^2|\Phi_0\rangle\right)
 -\frac{2}{9}|\langle \Phi_0|\phi_1+\phi_2+\phi_3|\Phi_1^{(0)}\rangle|^2\nonumber\\
 \!\!\!\!\!\!\!\!&&
 -{2/3}\left(\langle \Phi_1^{(0)}|\phi_1+\phi_2+\phi_3|\Phi_1^{(0)}\rangle
     \langle \Phi_0|\phi_1+\phi_2+\phi_3|\Phi_0\rangle
 -\langle \Phi_0|\phi_1+\phi_2+\phi_3|\Phi_0\rangle^2\right)\Big]
\nonumber\\
&=& 5.296 \ (2.710)~,
\nonumber\\
\widetilde{c}_1^{(0)}\Big|_{\rm magn}^{\rm 1st\ ord}
\!\!\!\!&=&\frac{2}{9}|\langle \Phi_0|\phi_1+\phi_2+\phi_3|\Phi_1^{(0)}\rangle|^2
= 0.050\ (0.048)~.
\nonumber
\end{eqnarray}
Together with the electric contributions
$c_1=14.161\ (4.660),\widetilde{c}_1= 0.3977\ (0.1778)$ and the
measure contributions $c_1=-1.813\ (-0.3597),\widetilde{c}_1=0.0119\ (0.0016)$, I obtain the total
value from first order perturbation theory
\begin{eqnarray}
&&c_1^{(0)}\Big|_{\rm tot}^{\rm 1st\ ord}
= 17.643\ (7.011)~,
\quad\quad
\widetilde{c}_1^{(0)}\Big|_{\rm tot}^{\rm 1st\ ord}
= 0.460\ (0.228)~.
\nonumber
\end{eqnarray}

%\subsubsection{Second order perturbation theory}

Furthermore, second order perturbation theory leads to a
change in the mass to to its virtual decay into two spin-2 particles or into one spin-2 and
one spin-3 particle.
Using ${\cal V}^{(\partial)}_\alpha$ in (\ref{calVp}) I obtain for the case of the virtual decay
into two spin-2 particles,
\begin{eqnarray}
\label{spin02nda}
\!\!\!\!\!\!\!\!\!\!\!\!\!\!\!\!\!\!\!
c_1^{(0)}\Bigg|^{\rm 2nd\ ord}_{2-2,\alpha,\alpha'}\!\!\!\!\!\!\!\!\!\!\!\!\!\!\!\!\!\!\!\!\!
&=&-\frac{\pi^2}{25}
\sum_{i_1,i_2}
\frac{1}
   {\mu_{i_1}^{(2)}+\mu_{i_2}^{(2)}-\mu_{1}^{(0)}}
  \Bigg[
        \langle \Phi_0||\widetilde{\mathbf{\cal Y}}^{(2)}_\alpha||\Phi_{i_1}^{(2)}\rangle
         \langle \Phi_{1}^{(0)}||{\mathbf{\cal Y}}^{(2)}_\alpha ||\Phi_{i_2}^{(2)}\rangle
       +\langle \Phi_{1}^{(0)}||\widetilde{\mathbf{\cal Y}}^{(2)}_\alpha||\Phi_{i_1}^{(2)}\rangle
         \langle \Phi_0||{\mathbf{\cal Y}}^{(2)}_\alpha ||\Phi_{i_2}^{(2)}\rangle
\Bigg]\!\!\times
 \nonumber\\
&&\quad\quad\quad\quad\quad\quad
\times\Bigg[
         \langle \Phi_{i_1}^{(2)}||\widetilde{\mathbf{\cal Y}}^{(2)}_{\alpha'}||\Phi_{0}\rangle
         \langle \Phi_{i_2}^{(2)}||{\mathbf{\cal Y}}^{(2)}_{\alpha'} ||\Phi_{1}^{(0)}\rangle
        +\langle \Phi_{i_1}^{(2)}||\widetilde{\mathbf{\cal Y}}^{(2)}_{\alpha'}||\Phi_{1}^{(0)}\rangle
         \langle \Phi_{i_2}^{(2)}||{\mathbf{\cal Y}}^{(2)}_{\alpha'} ||\Phi_{0}\rangle
 \nonumber\\
&&\quad\quad\quad\quad\quad\quad\quad\quad
        +\langle \Phi_{i_2}^{(2)}||\widetilde{\mathbf{\cal Y}}^{(2)}_{\alpha'}||\Phi_{0}\rangle
         \langle \Phi_{i_1}^{(2)}||{\mathbf{\cal Y}}^{(2)}_{\alpha'} ||\Phi_{1}^{(0)}\rangle
         +\langle \Phi_{i_2}^{(2)}||\widetilde{\mathbf{\cal Y}}^{(2)}_{\alpha'}||\Phi_{1}^{(0)}\rangle
         \langle \Phi_{i_1}^{(2)}||{\mathbf{\cal Y}}^{(2)}_{\alpha'} ||\Phi_{0}\rangle
\Bigg]~,
\\
\label{spin02ndb}
\widetilde{c}_1^{(0)}\Bigg|^{\rm 2nd\ ord}_{2-2,\alpha,\alpha'}
\!\!\!\!\!\!\!\!\!\!\!\!\!\!\!\!\! &=&\frac{1}{100}
\sum_{i_1,i_2}
\frac{1}
   {\mu_{i_1}^{(2)}+\mu_{i_2}^{(2)}-\mu_{1}^{(0)}}
 \Bigg[
       \langle \Phi_0||\widetilde{\mathbf{\cal Y}}^{(2)}_\alpha||\Phi_{i_1}^{(2)}\rangle
         \langle \Phi_{1}^{(0)}||{\mathbf{\cal Y}}^{(2)}_\alpha ||\Phi_{i_2}^{(2)}\rangle\times
\nonumber\\
&&\quad\quad\quad\quad\quad\quad\quad\quad\quad
 \times\Bigg(
        \langle \Phi_{i_1}^{(2)}||\widetilde{\mathbf{\cal Y}}^{(2)}_{\alpha'}||\Phi_{1}^{(0)}\rangle
      \langle \Phi_{i_2}^{(2)}||{\mathbf{\cal Y}}^{(2)}_{\alpha'}||\Phi_{0}\rangle
        +\langle \Phi_{i_2}^{(2)}||\widetilde{\mathbf{\cal Y}}^{(2)}_{\alpha'}||\Phi_{0}\rangle
         \langle \Phi_{i_1}^{(2)}||{\mathbf{\cal Y}}^{(2)}_{\alpha'} ||\Phi_{1}^{(0)}\rangle
\Bigg)
 \nonumber\\
&&\quad\quad\quad\quad\quad\quad
       +\langle \Phi_{1}^{(0)}||\widetilde{\mathbf{\cal Y}}^{(2)}_\alpha||\Phi_{i_1}^{(2)}\rangle
         \langle \Phi_0||{\mathbf{\cal Y}}^{(2)}_\alpha ||\Phi_{i_2}^{(2)}\rangle
 \Bigg(
         \langle \Phi_{i_2}^{(2)}||\widetilde{\mathbf{\cal Y}}^{(2)}_{\alpha'}||\Phi_{1}^{(0)}\rangle
  \langle \Phi_{i_1}^{(2)}||{\mathbf{\cal Y}}^{(2)}_{\alpha'} ||\Phi_{0}\rangle
 \nonumber\\
&&\quad\quad\quad\quad\quad\quad\quad\quad\quad\quad\quad\quad\quad\quad\quad\quad\quad\quad
\quad\quad\quad\quad\quad
         +\langle \Phi_{i_1}^{(2)}||\widetilde{\mathbf{\cal Y}}^{(2)}_{\alpha'}||\Phi_{0}\rangle
       \langle \Phi_{i_2}^{(2)}||{\mathbf{\cal Y}}^{(2)}_{\alpha'} ||\Phi_{1}^{(0)}\rangle
\Bigg)
\Bigg],
\end{eqnarray}
and similar expressions for the virtual decay into one spin-2 and one spin-3 particle.
The leading contribution to (\ref{spin02nda}) and (\ref{spin02ndb}) comes from
${\cal V}^{(\partial)}_{\rm magn}-{\cal V}^{(\partial)}_{\rm magn}$ with
$c_1=-3.019 \ (-2.5238),\widetilde{c}_1=0.0226\ (0.0194)$.
Together with the smaller contributions  $c_1=-0.9947 \ (-1.001),
\widetilde{c}_1=0.0055\ (0.0005)$ from
${\cal V}^{(\partial)}_{\rm magn}-{\cal V}^{(\partial)}_{\rm meas}$, the contributions
$c_1=-0.1189 \ (-0.0013),\widetilde{c}_1=0.0036\ (3 \times 10^{-6})$ from
 ${\cal V}^{(\partial)}_{\rm meas}-{\cal V}^{(\partial)}_{\rm meas}$,
 and the negligibly small contributions
 $c_1=-8.3 \times 10^{-5} \ (-2.4 \times 10^{-4}),
     \widetilde{c}_1=1.2 \times 10^{-7}\ (2.3 \times 10^{-7})$
 from ${\cal V}^{(\partial)}_{2-3}-{\cal V}^{(\partial)}_{2-3}$,
I obtain the total value from second order perturbation theory
\begin{eqnarray}
&&c_1^{(0)}\Big|_{\rm tot}^{\rm 2nd\ ord}
= -4.133\ (-2.626)~,
\quad\quad
\widetilde{c}_1^{(0)}\Big|_{\rm tot}^{\rm 2nd\ ord}
= 0.028\ ( 0.020)~.
\nonumber
\end{eqnarray}
First and second order perturbation theory give the results (up to $\lambda^2$)
\begin{eqnarray}
 E_1^{(0)+}(k)-E_{\rm vac}^{+} &=&
 \left[\ 2.270 + 13.510\lambda^2 + {\cal O}(\lambda^3)\right]\frac{g^{2/3}}{a}
 +0.488 \frac{a}{g^{2/3}} k^2 +{\cal O}((a^2 k^2)^2) ~,
 \\
 E_1^{(0)-}(k)-E_{\rm vac}^{-} &=&
 \left[\ 3.268
 + 4.385\lambda^2 + {\cal O}(\lambda^3)\right]\frac{g^{2/3}}{a} +
  0.248 \frac{a}{g^{2/3}} k^2 +{\cal O}((a^2 k^2)^2) ~.
\end{eqnarray}
for the energy spectrum of the interacting spin-0 glueball
for the $(+)$ and $(-)$ boundary conditions, respectively.
The results are summarized in Table 3
\footnote{
%-------------------------------------------------------------------------------------------------------
I would like to comment here, that, using only nearest neighbor interactions ($N=1$ in (\ref{1stdisc}) and
(\ref{2nddisc}) instead of $N\rightarrow\infty$), it would lead to the same $\widetilde{c}_1$, but
a $(\pi^2/6)\simeq 1.64$ times smaller ${c_1^{({\rm 1st})}}|_{N=1}=10.726 (4.262)$
and a $(2\pi^2/3)\simeq 6.58$ times smaller ${c_1^{({\rm 2nd})}}|_{N=1}=-0.628 (-0.399)$ and hence to
a $25\% (12\%)$ smaller $c_1|_{N=1}= 10.097 (3.863)$. Similarly for the vacuum (Table 2),
it would lead to a $38\%$ smaller $c_0|_{N=1}= 18.438 (9.041)$.
%---------------------------------------------------------------------------------------------------------
}.
\newline
\newline
$\begin{array}{|c|c||c|c|c|c||c|c|c|c|}
{\rm spin}-0  & \mu_1 & c_1^{({\rm 1st})} & c_1^{({\rm 2nd})} &
 c_1 & c_1/\mu_1
& \widetilde{c}_1^{({\rm 1st})} & \widetilde{c}_1^{({\rm 2nd})}&
\widetilde{c}_1  & 1/(2\mu_1)\\  \hline\hline
(+)  & 2.270 & 17.643 & -4.133 & 13.510 & 5.953 & 0.460 & 0.028 & 0.488 & 0.220 \\ \hline
(-)  & 3.268 &\ 7.011 & -2.626 &\ 4.385 & 1.342 & 0.228 & 0.020 & 0.248 & 0.153 \\ \hline
\end{array}$
\newline
\newline
Table 3: Results for the interacting spin-0 glueball for(+) and (-) b.c.
The numerical errors are smaller than the last digits in the numbers shown.

\subsection{Discussion of the results}

First I would like to comment on the relation between
the glueball mass and coupling constant renormalisation in the IR.
Consider the physical mass
\begin{equation}
\label{flow}
M = \frac{g_0^{2/3}}{a}\left[\mu +c g_0^{-4/3}\right]~.
\end{equation}
Demanding its independence of box size $a$,
one obtains
\begin{eqnarray}
\gamma(g_0) \equiv a {d\over da} g_0(a) =
  \frac{3}{2} g_0\frac{\mu+c g_0^{-4/3}}
     {\mu-c g_0^{-4/3}}
\nonumber
\end{eqnarray}
which vanishes for the two cases,
$g_0 = 0 $ or
$g_0^{4/3}=-c/\mu$.
The first solution corresponds to the perturbative fixed point, and the second,
if it exists $(c<0)$, to an infrared fixed point.
My result for $ c_1^{(0)}/\mu_1^{(0)}=5.95  (1.34) $
suggests, that
no infrared fixed points exist,
in accordance with the corresponding result of Wilsonian lattice QCD\footnote{
%-------------------------------------------------------------------------------
In comparison, the $SU(2)$ result from strong coupling on the lattice \cite{Muenster},\cite{Creutz}:
$ a M =4\log(g_0^2)+ O(g_0^{-2})
\rightarrow
\gamma(g_0) = \frac{1}{2} g_0\log(g_0^2)+ ... $
does not contain infrared fixed points.
%-------------------------------------------------------------------------------
}.
Solving the above equation (\ref{flow}) for positive $(c>0)$ I obtain
\begin{eqnarray}
g_0^{2/3}(M a) =\frac{M a}{2\mu}
     +\sqrt{\left(\frac{M a}{2\mu}\right)^2
     -\frac{c}{\mu}}~,\quad
     \  a >  a_{c} :=  2\sqrt{c \mu}/M
\label{largebox}
\end{eqnarray}
with the physical glueball mass $M$.
The minimal lattice sizes are $M a_{c}=11.08\ (7.57)$, corresponding to a
critical coupling $g^2_{0}|_{c}= 14.52\ (1.55)$.
Taking a typical physical glueball mass of
$M \sim 1.6\ {\rm GeV}$ \cite{Teper}, I obtain
\begin{eqnarray}
{\rm for}\ M \sim 1.6\ {\rm GeV}:\quad a_{c} \sim 1.4\ {\rm fm}\ (0.9\ {\rm fm})~,
\nonumber
\end{eqnarray}
which seems reasonable. The dependence of the results on the boundary conditions imposed,
$(+)$ or $(-)$, might be seen as a prescription dependence. Of course, it will be much more
effective to consider mass ratios, as soon as, for example, the spin-2 glueball will be calculated.
It would also be interesting, to connect the behaviour of the glueball spectrum and the
bare coupling constant (\ref{largebox}),
obtained for boxes of large size a, with those obtained for small boxes
(see \cite{Luescher},\cite{Luescher and Muenster} and footnote \ref{smallbox}),
in order to get information about the intermediate region, including the possibility
of the occurrence of phase transitions.

Furthermore, I would like to remark, that Lorentz invariance imposes the following condition
on the coefficients $\widetilde{c}_i^{(S)}$ in (\ref{spin0sp}):
\begin{eqnarray}
  E=\sqrt{m^2+k^2}\simeq m + \frac{1}{2 m}\ k^2
        \quad \rightarrow \quad  \widetilde{c}_i^{(S)} = 1/(2\mu_i^{(S)})\nonumber~.
\end{eqnarray}
Comparison of the last two columns of Table 3 show that my result does not satisfy this
requirement by a factor of about 2.
Of course, the glueball excitation carrying non-relativistic spin-0 considered here
as a first step, does not correspond to a relativistic particle. Hence states of total
angular momentum $J=S+L$, containing spin and orbital angular momentum, similar to the
quark states in the Dirac wave-function, should be considered, e.g. the $J=0$ state
(using spherically symmetric granulation)
\begin{eqnarray}
|J=0, k\rangle \!\!\!\!&\sim &\!\!\!\! \alpha_1^{(0)}\sum_{\mathbf{n}}
 j_0(k r)\left[|\Phi_1^{(0)}\rangle_{\mathbf{n}}\bigotimes_{{\mathbf{m}}\neq {\mathbf{n}}}
 |\Phi_0\rangle_{\mathbf{m}}\right]
+\sum_{S,i}^{\rm stable}\alpha_i^{(S)}\sum_{\mathbf{n}} j_S(k r)\sum_{M}
Y_{SM}(\theta,\phi)\left[|\Phi_{i,M}^{(S)}\rangle_{\mathbf{n}}
\bigotimes_{{\mathbf{m}}\neq {\mathbf{n}}}
 |\Phi_0\rangle_{\mathbf{m}}\right]
\nonumber~,
\end{eqnarray}
where the sum is over all excitations $\mu_i^{(S)} < \mu_{\rm th}$ , underlined in Table 1,
which are stable at tree-level.
For simplicity, I have considered in this work only the spin-0 excitation $\mu_1^{(0)+}$ ($\mu_1^{(0)-}$),
but of course ,
also the lowest spin-2 excitations $\mu_1^{(2)+},\mu_2^{(2)+}$ ($\mu_1^{(2)-}$)
and the lowest spin-4 excitation $\mu_1^{(4)+}$ ($\mu_1^{(4)-}$) have to be included.
Most important will certainly be the inclusion of the spin-2 state $\mu_1^{(2)+}$ ($\mu_1^{(2)-}$),
which is lower in energy than the spin-0 state considered in this work.
The necessary extension of the calculation to spin-2 and spin-4 states
and the inclusion of orbital angular momentum of the lowest excitations
clearly goes beyond the scope of this work.

%%%%%%%%%%%%%%%%%%%%%%%%%%%%%%%%%%%%%%%%%%%%%%%%%%%%%%%%%%%%%%%%%%%%%%%%%%%%%%%%%%%%%%%%%%%%%%%%

\section{Conclusions}

It has been shown in this work, how a gauge invariant formulation of Yang-Mills theory on a
3-dimensional spatial lattice can be obtained by  replacing integrals by sums
and spatial derivatives by differences.
This has been achieved by using the symmetric gauge $\epsilon_{ijk}A_{jk}=0$ \cite{KP1}\cite{KMPR},
and constructing the corresponding physical quantum Hamiltonian of $SU(2)$ Yang-Mills theory
according to the general scheme given by Christ and Lee \cite{Christ and Lee}.
In contrast to the Coulomb gauge formulation, very suitable for the description of the
high energy sector of the theory, the symmetric gauge quantum Hamiltonian, obtained here,
is very suitable for the IR sector, since it can be expanded in the number of spatial derivatives.
The "derivative-free" part of the Hamiltonian is just the sum of Hamiltonians of Yang-Mills quantum mechanics
of constant fields for each granula (here a box of size a), with a purely discrete spectrum
("free glueballs"), and the terms of higher and higher number
of spatial derivatives describing interactions between the constant fields of different granulas.
This expansion has been carried out here explicitly and shown to be equivalent to a strong coupling expansion
in $\lambda=g^{-2/3}$ for large box sizes $a$. It is the analogon
to the weak coupling expansion in $g^{2/3}$ by L\"uscher and M\"unster \cite{Luescher}
\cite{Luescher and Muenster}, applicable for small boxes.
Using the very accurate results of Yang-Mills quantum mechanics of constant fields in a box,
obtained with the variational method in earlier work \cite{pavel},
the energy spectrum of weakly interacting glueballs can be calculated systematically and with high accuracy,
using perturbation theory in $\lambda$.
This offers a useful alternative to lattice calculations based on the Wilson-loop,
including the corresponding analytic strong coupling expansions by Kogut, Sinclair, and Susskind \cite{Kogut}
and M\"unster \cite{Muenster}.
My result for the mass of the interacting spin-0 glueball up to $\lambda^2$, as a first step, confirms
the result of Wilsonian lattice QCD, that no infrared fixed points exist.
Problems are a.o. the question of Lorentz invariance of the glueball spectrum.

\section*{Acknoledgements}

I would like to thank A. Dorokhov, V. Ponomarev,  J. Wambach, and W. Weise
for their interest and support.
%%%%%%%%%%%%%%%%%%%%%%%%%%%%%%%%%%%%%%%%%%%%%%%%%%%%%%%%%%%%%%%%%%%%%%%%%%%%%%%%%%%%%%%%%%%%%%

\end{document}